\documentclass[11pt] {article}

\usepackage{natbib}
\bibpunct{[}{]}{;}{n}{,}{,}

\usepackage{amsmath}
\usepackage{graphicx,latexsym, amssymb,epsfig}
 
\usepackage{float} 
\date{}

\newcommand{\HRule}[1]{\rule{\linewidth}{#1mm}}

\title{
\Large  {\bf The Merger of the Schwarzschild metric and the Robertson-Walker metric}\\
\normalsize {Bethmini Senevirathne$^1$ and Nalin de Silva$^2$\\
$^1$Department of Physics and Astronomy, University of Glasgow, UK\\
$^2$Department of Mathematics, University of Kelaniya, Sri Lanka}
}

\begin{document}
\maketitle
\vspace*{-2cm}

\HRule{0.1}

\section{Introduction}

The Schwarzschild metric that represents the space time outside a spherically symmetric object is usually derived with the Lorentz space time in the background. In other words in deriving the Schwarzschild metric it is assumed that the metric becomes the Lorentz metric at large distances.\\

However, we live in a universe generally believed to be represented by the Robertson-Walker metric and the assumption that the space time due to a spherically symmetric object becomes the Minkowskian space time at large distances cannot be taken to be correct. The space time due to a spherically symmetric object should merge with the Robertson-Walker metric at some distance and the imposition of the boundary conditions would give a metric with properties different from the Schwarzschild metric.\\

In this paper we derive a metric due to a spherically symmetric object, that merges with the Robertson-Walker metric at coordinate $r$ to be specified. The derivation is similar to that given in Adler, Bazin and Schiffer (1965) in the case of a spherically symmetric object that gives the Lorentz metric at large distances.

\section{Merger of the two metrics}
Let 
\begin{eqnarray*}
ds^2&=&e^{\nu (r)}c^2dt^2-e^{\lambda(r)}dr^2-r^2(d\theta^2+\sin^2\theta d\phi ^2), \qquad r \leq a\\
ds^2&=&c^2dT^2-\left(\frac{R(T)}{R^{\ast}}\right)^2\left(\frac{du^2}{1-\frac{ku^2}{(R^{\ast})^2}}+u^2(d\Theta^2+\sin^2\Theta d\Phi^2)\right), \qquad u \geq b
\end{eqnarray*}
be the metrics due to a spherically symmetric object, where $u$, $\Theta$, $\Phi$ are the comoving coordinates  in the Robertson-Walker  metric. The two metrics are assumed to merge at $r=a$  and  $u=b$.\\

It can be assumed without loss of generality that $\theta=\Theta$  and $\phi=\Phi$   as the angular coordinates do not change when the metrics are merged.\\

Using the calculation given in Adler, Bazin and Schiffer we obtain $\nu'+\lambda'=0$   and hence $\nu+\lambda=$constant = $A$.\\

The boundary condition that the metric becomes Lorentzian at large distances makes $\lambda=-\nu$, and $A = 0$ in the derivation of Adler, Bazin and Schiffer.  However in our derivation we take $A\neq 0$.\\

Thus we find, after going through a calculation similar to that given in Adler, Bazin and Schiffer,   
that 
\begin{equation*}
e^{\lambda}=\frac{C}{1-\frac{2m}{r}} \quad \text{ and }\quad e^{\nu}=B\left(1-\frac{2m}{r}\right), \quad \text{ where } \quad BC=e^A, m=\frac{GM}{c^2},  
\end{equation*}
$M$  being the mass of the spherically symmetric object, $G$ and  $c$  having the usual meanings. It is also seen that $B$ and $C$ are both positive. \\

Hence 
\begin{eqnarray*}
ds^2&=&B\left(1-\frac{2m}{r}\right)c^2dt^2-\left(\frac{dr^2}{1-\frac{2m}{r}}+r^2(d\theta^2+\sin^2d\phi^2)\right), \qquad r\leq a,\\
ds^2&=&c^2dT^2-\left(\frac{R}{R^{\ast}}\right)^2\left(\frac{du^2}{1-\frac{ku^2}{(R^{\ast})^2}}+u^2(d\theta^2+\sin^2\theta d\phi^2)\right), \qquad u \geq b.
\end{eqnarray*}
We use the following boundary conditions at $r = a$ and $u = b$. 
\begin{eqnarray}
\sqrt{B\left(1-\frac{2m}{r}\right)}c\delta t &=&c \delta T\\
\frac{\delta r}{\sqrt{C(1-\frac{2m}{r})}}&=&\left(\frac{R}{R^{\ast}}\right)\frac{\delta u}{\sqrt{1-\frac{ku^2}{(R^{\ast})^2}}}\\
r&=&\left(\frac{R}{R^{\ast}}\right)u\\
\sqrt{C\left(1-\frac{2m}{r}\right)}\frac{d}{dr}\left\{B\left(1-\frac{2m}{r}\right)\right\}c\delta t&=&\left(\frac{R^{\ast}}{R}\right)\sqrt{1-\frac{ku^2}{R^2}}\frac{d}{du}(1)c\delta T=0
\end{eqnarray}[Use the boundary condition $\dfrac{1}{\sqrt{g_{11}}}\dfrac{\partial}{\partial x_1}(g_{00})\delta t-\dfrac{1}{\sqrt{G_{11}}}\dfrac{\partial}{\partial X_1}(G_{00})\delta T$, with respect to the metrics $$ds^2=g_{00}(dx^0)^2+g_{11}(dx^1)^2+g_{22}(dx^2)^2+g_{33}(dx^3)^2 \text{ and }$$   $$ds^2=G_{00}(dx^0)^2+G_{11}(dx^1)^2+G_{22}(dx^2)^2+G_{33}(dx^3)^2.$$

This is a new boundary condition derived by Wimaladharma and de Silva(2008) based on  
$\left(\frac{\partial \phi}{\partial r}\right)_1=\left(\frac{\partial \phi}{\partial r}\right)_2$ in Newtonian Physics. ]\\

From (4) we have $\delta t=0$\\
and from (3) 
\begin{equation}
\delta r=\frac{u}{R}\delta R+\frac{R}{R^{\ast}}\delta u
\end{equation}
From  (1)  $\delta t =0 \implies \delta T=0$,\\
and since $\dfrac{dR}{dt}\neq 0$ $\delta T=0 \implies \delta R=0$.\\
$\therefore$ from   (5)  
\begin{equation}
\delta r=\frac{R}{R^{\ast}}\delta u
\end{equation}
\begin{equation*}
\text{Now (2) and (6)  } \implies \sqrt{C\left(1-\frac{2m}{r}\right)} =\sqrt{1-\frac{ku^2}{R^2}} 
\end{equation*}
\begin{equation}
\text{i.e. }C-\frac{2mC}{r}=1-\frac{ku^2}{R^2}
\end{equation}
Since $r=\dfrac{Ru}{R^{\ast}}$
\begin{equation}
(7) \implies \frac{kr^3(R^{\ast})^2}{R^4}+r(C-1)-2mC=0
\end{equation}
$\therefore $ at the boundary
$$a=\frac{2mC}{C-1}\qquad \text{ and }\qquad b=\frac{2mCR^{\ast}}{(C-1)R}\qquad \text{ if }k=0.$$

Now since $C>0$, $a$ is positive if $C>1$. Then $\dfrac{C}{C-1}>1$ and $a>2m$.

Then the merger is at a ``coordinate distance'' greater than the Schwarzschild radius of the object.\\

If $r_0$ is the ``radius'' of the spherically symmetric object, then $a \gtrless r_0$ if $\dfrac{2mC}{C-1} \gtrless r_0$.\\

Depending on $m$, $r_0$ and $C$ there is a possibility that $a$ the ``coordinate distance'' at the merger is less than $r_0$ the ``radius'' of the object.\\

When $k= 1$, it can be shown that there is positive real root of (8) and denoting it by $a$ we can obtain a similar result. When $k= - 1$,there is either no positive solution implying that there is no merger, or two positive solutions with the lesser value corresponding to the merger.

\section{Discussion}
The ``Schwarzschild'' metric due to a spherically symmetric object merges with the Robertson-Walker metric at a ``coordinate distance'' $\dfrac{2mC}{C-1}$ when $k=0$. For certain values of the mass and the ``radius'' of the object, this distance may be greater than the ``radius''. On such occasions the body is separated from the rest of the universe and the object can be ``seen'' by distant observers. However, when the merger is at a ``distance'' less than the ``radius'' of the object, the body is not separated from the rest of the universe and a portion of the body may not be ``seen'' by the distant observers, though gravitationally it interacts with the rest of the universe. The ``unseen'' portion of the object could be thought to be constituted of dark matter.\\

A similar result could be obtained if $k=\pm 1$, under certain conditions.

\section*{References}
\begin{enumerate}
\item[1.] Adler R., Bazin M., and Schiffer M (1965), Introduction to General Relativity, McGraw-Hill Book Company.
\item[2.] Wimaladharma N.A.S.N., and de Silva Nalin (2008), A metric which represents a sphere of constant uniform density comprising electrically counterpoised dust, Annual Research Symposium, University of Kelaniya, Sri Lanka.
\end{enumerate}
\end{document}